\documentclass{IEEEtran}
\usepackage[utf8]{inputenc}
\usepackage[T1]{fontenc}
\usepackage{comment}
\usepackage{amsmath}
\usepackage{amsfonts}
\usepackage[english]{babel}
\usepackage{amsthm}
\usepackage{amssymb}
\usepackage{graphicx}
\usepackage{latexsym}
\usepackage{algorithm}
\usepackage{algpseudocode}
\usepackage{lipsum}
\usepackage{enumitem}
\usepackage{tikz, pgfplots}
\usepackage{eso-pic,xcolor}
\usepackage{overpic, scalerel}
\pgfplotsset{compat=1.18}
\IEEEoverridecommandlockouts
\usepackage{subcaption}
\usepackage{multicol}
\usepackage{hyperref}
\usepackage[hyphenbreaks]{breakurl}
\usepackage{cleveref}
\usepackage{multirow}
\usepackage{algorithm}
\usepackage{algpseudocode}
\newcommand{\customref}[2]{\hyperref[#1]{#2}}
\usepackage[normalem]{ulem} 

\newcommand{\ulink}{\scaleto{DL}{4.0pt}}

\newcommand{\risg}{\scaleto{g}{4.0pt}}

\newcommand{\transpose}{\mkern-1mu\scaleto{\mathrm{T}}{3.5pt}}

\newcommand{\herme}{\mkern-1mu\scaleto{\mathrm{H}}{4.0pt}}

\newcommand{\ul}{\mkern-1mu\scaleto{\mathrm{(1)}}{5.5pt}}
\newcommand{\dl}{\mkern-1mu\scaleto{\mathrm{(2)}}{5.5pt}}
\newcommand{\kl}{\mkern-1mu\scaleto{\mathrm{*}}{4.5pt}}
\newcommand{\cas}{\mkern-1mu\scaleto{\mathrm{cas}}{3.5pt}}

\newcommand{\pil}{u}

\definecolor{customGreen}{rgb}{0, 0.502, 0}

\newcommand{\e}{\mathbb{E}}

\newcommand{\RIS}{\scaleto{H}{4.5pt}}
\newcommand{\RISg}{\scaleto{g}{4.0pt}}

\newcommand\scalemath[2]{\scalebox{#1}{\mbox{\ensuremath{\displaystyle #2}}}}

\usepackage[figurename=Fig.]{caption}
\newcommand{\captionwidth}{\captionsetup{width=\linewidth}}
\usepackage{physics}

\title{Low-Complexity Channel Estimation Protocol for Non-Diagonal RIS-Assisted Communications}

\author{{Mostafa Samy,~\IEEEmembership{Graduate Student Member,~IEEE}, Hayder Al-Hraishawi,~\IEEEmembership{Senior Member,~IEEE},\\ Abuzar B. M. Adam,~\IEEEmembership{Member,~IEEE}, Madyan Alsenwi,~\IEEEmembership{Member,~IEEE}\\ Symeon Chatzinotas,~\IEEEmembership{Fellow,~IEEE}, and Björn Otteresten,~\IEEEmembership{Fellow,~IEEE}\vspace{-10mm}}\\
\thanks{M. Samy, A. B. M. Adam, M. Alsenwi, and B. Otteresten are with the Interdisciplinary Centre for Security, Reliability and Trust (SnT), University of Luxembourg, Luxembourg. \\
H. Al-Hraishawi is with the Department of Electrical Engineering, University of South Florida, Tampa, FL 33620 USA.\\
S. Chatzinotas is with SnT, University of Luxembourg, 1855 Luxembourg City, Luxembourg and with College of Electronics \& Information, Kyung Hee University, Yongin-si, 17104, Korea.\\
Corresponding author: \emph{Mostafa Samy (mostafa.samy@uni.lu)}.}
\thanks{This work was supported in part by the Luxembourg  National Research Fund (FNR) through the AFR Project CEP-MBD-CRIS  grant reference 17974844, and in part by the FNR grant reference INTER/MOBILITY/2023/IS/18014377/MCR. }}

\begin{document}
\pagenumbering{arabic}
\maketitle
\bstctlcite{IEEEexample:BSTcontrol}
\begin{abstract} 
Non-diagonal reconfigurable intelligent surfaces (RISs) offer enhanced wireless signal manipulation over conventional RIS by enabling the incident
signal on any of its $M$ elements to be reflected from another element via an $M \times M$ switch array. To fully exploit this flexible configuration, the acquisition of individual channel state
information (CSI) is essential. However, due to the passive nature of the RIS, cascaded channel estimation is performed, as the RIS itself lacks signal processing capabilities. This entails estimating the CSI for the $M \times M$ switch array cascaded channels, resulting in estimating $M^2$ coefficients, to identify the optimal configuration of the non-diagonal RIS that maximizes the channel gain. 
In this paper, we propose a low-complexity channel estimation protocol that substantially reduces the estimation overhead from the exhaustive $M^2$ coefficients to only $3M$ coefficients for both single-input single-output (SISO) and multiple-input single-output (MISO) systems. Specifically, a three-stage pilot-based protocol is proposed to estimate scaled versions of the user-RIS and RIS-base-station (BS) channels in the first two stages using the least square (LS) estimator and the commonly used ON/OFF protocol for conventional RIS. These scaled estimates enable the optimization of the switch array. In the third stage, the cascaded user-RIS-BS channels are estimated to enable efficient beamforming optimization. 
Complexity analysis shows that our proposed protocol significantly reduces the BS estimation complexity from $\mathcal{O}(NM^2)$ to $\mathcal{O}(NM)$, where $N$ is the number of BS antennas. This complexity is similar to the ON/OFF-based LS estimation for conventional diagonal RIS.
\end{abstract}

\section{Introduction}
Reconfigurable intelligent surfaces (RISs) are regarded as a key enabler for 6G wireless networks, offering the ability to manipulate wireless propagation channels and enhance spectrum and energy efficiency using low-cost hardware \cite{Wang2023}. 
Composed of a large array of reconfigurable and energy-efficient passive elements, RIS can dynamically control the phase and amplitude of incident electromagnetic waves, creating a programmable wireless environment that optimizes communication performance \cite{Kisseleff2020}. These compelling features have positioned RIS as a focal point of research  for enhancing spectral and energy efficiency gains in wireless communications \cite{Mostafaoj}. Particularly, RISs have demonstrated effectiveness in improving various aspects of wireless systems, including full-duplex communication, cooperative relaying, physical layer security, and integrated sensing and communications (ISAC) capabilities \cite{Ahmed2024survey,Zuo2023}.

Structurally, an RIS consists of multiple scattering elements integrated with a reconfigurable impedance network. Traditional single-connected RIS architectures mainly allow only phase tuning of incident waves, thereby limiting their ability  to effectively manipulate the propagation environment.  To overcome these limitations, RIS 2.0 or beyond diagonal RIS (BD-RIS) \cite{bruno2,bruno3} has been introduced as an advanced approach to expand RIS functionality. In particular, the non-diagonal RIS architecture proposed in \cite{non_diagonal} enables signals impinging on one element to be reflected from another element after a phase-shift adjustment, achieved through a simple switch array. Hence, with the switch array in the non-diagonal RIS, the scattering matrix is no longer modeled as symmetric as in traditional RISs, providing greater flexibility to further enhance system performance. Compared to the fully- and group-connected architectures \cite{bruno2}, non-diagonal RIS structures minimize the number of required phase shifters by utilizing a switch array to connect between elements instead of phase-shifters.
Specifically, this configuration capitalizes on recent advancements in energy-efficient switching technologies \cite{masoro_switch}, offering advantages over methods that depend on tunable impedance components with high-power consumption.

Despite the promising potential of non-diagonal RIS, the critical challenge of acquiring channel state information (CSI) has not been adequately addressed. From a practical standpoint, the achievable performance gains of non-diagonal RIS-assisted communication systems depend on obtaining the RIS individual CSI to fully leverage the unique interconnection capabilities between non-diagonal RIS elements. 
Consequently, for a passive RIS, the base-station (BS) needs to perform cascaded channel estimation for $M^2$ channels of the switch array to identify the optimal configuration that maximizes channel gains.

Motivated by this observation, and arguably, as the first work in this area, we propose a low-complexity channel estimation scheme for non-diagonal RIS-assisted communication systems, considering both single-input single-output (SISO) and multiple-input single-output (MISO) systems. Our approach integrates the general ON/OFF protocol for conventional RIS channel estimation with the unique capabilities of the non-diagonal RIS switch array and the least squares (LS) estimator. This combination enables the BS to acquire a scaled version of the individual CSI required to determine the optimal non-diagonal RIS elements mapping. Subsequently, the cascaded channel for this optimal configuration is estimated to enable both active and passive beamforming optimization. More importantly, our method eliminates the need for active sources at the RIS, aligning with the core concept of passive RIS design, thereby enhancing its practicality for real-world deployment.

\subsection{Notation}
Vectors and matrices are denoted by boldface lower and upper case symbols, respectively. $(\cdot)^{\transpose}$ and $(\cdot)^{\herme}$ represents the operation of transpose and Hermitian, respectively. The notation $||\cdot||$ denotes the Euclidean norm, while ${\left\|  \cdot  \right\|_F}$  denotes the Frobenius norm.
The notation $[\mathbf{a}]_m$ denotes the $m$-th element of the vector $\mathbf{a}$, and $[\mathbf{A}]_{n,m}$ denotes the $(n, m)$-th element of the matrix $\mathbf{A}$. The notations $|\mathbf{a}|$ and $\angle \mathbf{a}$ represent the amplitude and phase of the complex vector $\mathbf{a}$, respectively.
The notation $ X \sim \mathcal {CN} (\mu_x,\sigma^2_x)$ denotes that random variable $X$ is circularly symmetric complex Gaussian distributed with $\mu_X$ mean and $\sigma^2_X$ variance.
$\mathbb E\{\cdot\}$ is the expectation operator. The notation $\otimes$ denotes the Kronecker product.
Further, $\operatorname{diag}\qty{a_1,\dots,a_M}$
denotes a diagonal matrix with diagonal elements $\qty(a_1,\dots,a_M)$. The notation $M!$ refers to the factorial of a positive integer $M$, defined as $M! = M \times (M-1) \times \dots \times 1$.

\begin{figure}
    \centering
    \includegraphics[width=3.3in]{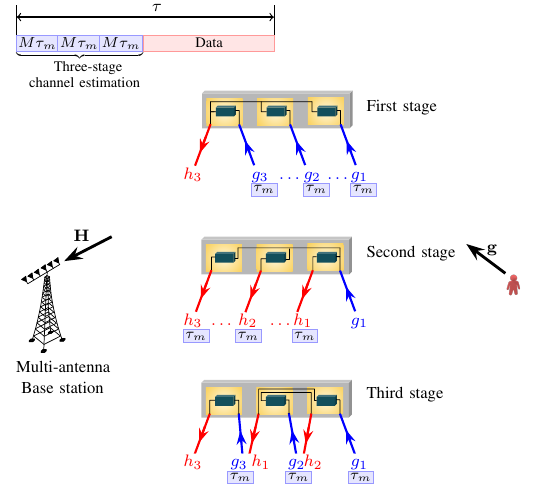}
\caption{\footnotesize  Illustration of the proposed three-stage channel estimation protocol. In the first stage, the switch array is employed to enable the incident signals on all elements to be reflected from a single element, while the ON/OFF protocol ensures that this reflection is performed orthogonally, one-by-one with each element having a duration of $\tau_{m}$. Similarly, in the second stage, the incident signal from a single element is reflected from all elements. The first and second stages enable the acquisition of scaled versions of $\mathbf{g}$ and $\mathbf{H}$, which are  sufficient to determine the optimal configuration of the switch array. In the third stage, the cascaded channel is estimated for this optimal configuration.}
       \vspace{-2.6mm}
          \label{sysmo}
\end{figure}
\section{System Model}\label{sec:sys_mod}
We consider a non-diagonal RIS-aided communication system that consists of a BS equipped with $N$ antennas, a non-diagonal RIS having $M$ passive reflecting elements, and a single-antenna user.
We assume that the direct link between the BS and user is obstructed, making communication feasible only through the non-diagonal RIS.
The non-diagonal RIS architecture enables a signal impinging on one element to be reflected by another element after a phase shift adjustment. This functionality is mathematically represented by an $M \times M$ phase shift matrix, which contains $M$ non-zero elements. The positions of these non-zero entries in the matrix are dictated by the mapping between the incident and reflected signals, effectively capturing the inter-element reflection behavior of the non-diagonal RIS.
For instance, the phase shift matrix for a non-diagonal RIS with three elements can be expressed as:
\begin{align}
\mathbf{\Theta}=\beta
\begin{bmatrix}
0 & e^{j\theta_{1,2}} & 0 \\
0 & 0 & e^{j\theta_{2,3}}  \\
e^{j\theta_{3,1}} & 0 & 0
\end{bmatrix},
\end{align}
where $\beta$ represents the amplitude coefficient, which is set to $1$ to achieve full reflection, and $\theta_{m,m'}$ denotes the phase shift applied by element $m$, for $m\in\qty{1,...,M}$ and $m'\in\qty{1,...,M}$.
Moreover, the non-diagonal matrix can be expressed in terms of the diagonal one as follows \cite{non_diagonal}
\begin{equation}\label{alternative}
\mathbf{\Theta} = \mathbf{J}_t \mathbf{\Phi}\mathbf{J}_r, 
\end{equation}
where $\mathbf{\Phi}=\operatorname{diag}\{e^{j\theta_1}, e^{j\theta_2}, \ldots, e^{j\theta_M}\}$, $\mathbf{J}_t\in\mathbb{C}^{M \times M}$ and $\mathbf{J}_r\in\mathbb{C}^{M \times M}$ represent the transmission and reflection permutation matrices, respectively. These permutation matrices define the mapping of incident and reflected signals through the non-diagonal RIS switch array, thereby enabling the realization of the non-diagonal structure of $\mathbf{\Theta}$ \cite{non_diagonal}. For example, the number of possible configurations for an $M$ elements non-diagonal RIS is $M!$ as shown in Table \ref{t1}. 
\begin{table*}[t]
\centering
\caption{Example of possible configurations for non-diagonal RIS with $M=3$ elements, showing $M^2 = 9$ cascaded channels.}
\renewcommand{\arraystretch}{1.3}  
\begin{tabular}{|c|c|c|c|c|c|c|}
\hline
\multirow{2}{*}{\emph{Incident}} & \multicolumn{6}{c|}{\emph{Possible configurations: 3! = 6}} \\
\cline{2-7}
 & \emph{Configuration 1} & \emph{Configuration 2} & \emph{Configuration 3} & \emph{Configuration 4} & \emph{Configuration 5} & \emph{Configuration 6} \\
\hline
\emph{Incident 1} & \textbf{reflected from 1} & \textbf{reflected from 2} & \textbf{reflected from 3} & reflected from 1 & reflected from 2 & reflected from 3 \\
\emph{Incident 2} & \textbf{reflected from 2} & \textbf{reflected from 1} & reflected from 2 & \textbf{reflected from 3} & reflected from 3 & reflected from 1 \\
\emph{Incident 3} & \textbf{reflected from 3} & reflected from 3 & \textbf{reflected from 1} & \textbf{reflected from 2} & reflected from 1 & reflected from 2 \\
\hline
\end{tabular}
\label{t1}
\end{table*}

\subsection{Channel Model}

We consider a flat quasi-static Rayleigh block fading model, where the channel for each link remains static during a coherence interval of $\tau$ seconds. In a time-division duplex (TDD) system, the BS performs channel estimation using uplink pilot signal transmitted by the user via the non-diagonal RIS and leverages channel reciprocity to obtain the downlink channel. The channel from the BS to the non-diagonal RIS is represented by $\mathbf{H} \in \mathbb{C}^{N \times M}$, where each entry $[\mathbf{H}]_{n,m}$ is independently and identically distributed (i.i.d.) following a complex Gaussian distribution, $[\mathbf{H}]_{n,m} \sim \mathcal{CN}(0, \zeta_H)$. Here, $\zeta_H = d_{\RIS}^{-\eta_{\RIS}}$ accounts for the path loss, where $d_{\RIS}$ is the distance between the BS and the non-diagonal RIS, and $\eta_{\RIS}$ denotes the path-loss exponent.
The non-diagonal RIS-to-user channel is represented by $\mathbf{g} \in \mathbb{C}^{M \times 1}$, where each entry $[\mathbf{g}]_m$ is i.i.d. following a complex Gaussian distribution, $[\mathbf{g}]_m \sim \mathcal{CN}(0, \zeta_g)$. Here, $\zeta_g = d{\risg}^{-\eta{\risg}}$ captures the path loss, with $d_{\risg}$ denoting the distance between the non-diagonal RIS and the user, and $\eta_{\risg}$ representing the associated path-loss exponent.
\subsection{Signal Model}\label{downl}
This subsection presents the downlink signal model for the non-diagonal RIS-assisted communication system, focusing on the key factors that influence the design of both active and passive beamforming. Building on this foundation, we introduce a channel estimation protocol for the non-diagonal RIS architecture. The downlink received signal at the user can be expressed as:
%
%
\begin{align}\label{DL_signal}
y^{\ulink} = 
\qty(\mathbf{H}\mathbf{\Theta}
\mathbf{g})^{\transpose}\mathbf{w}s + z_{u},
\end{align}
where $s$  denotes the transmit signal at the BS satisfying $\mathbb E\qty[|s|^2]=1$, $\mathbf{w} \in \mathbb{C}^{N \times 1}$ represents the active beamforming vector at the BS satisfying $\|\mathbf{w}\| = 1$, and $z_{u}$ is the additive white Gaussian noise (AWGN) at the user; $z_{u} \sim \mathcal {CN} (0,\sigma^2_z)$. Substituting the non-diagonal matrix $\mathbf{\Theta}$ in \eqref{alternative} into \eqref{DL_signal} gives the received signal as follows 
\begin{align}\label{equ1}
y^{\ulink} = 
\mathbf{g}^{\transpose}
\mathbf{J}_r^{\transpose}
\mathbf{\Phi}
\mathbf{J}_t^{\transpose}
\mathbf{H}^{\transpose}\mathbf{w}s + z_{u}.
\end{align}
Furthermore, \eqref{equ1} can be rewritten as follows 
\begin{align}
y^{\ulink} = 
\mathbf{q}^{\transpose} 
\operatorname{diag}\{\mathbf{g}^{\transpose}\mathbf{J}_r^{\transpose}\} \mathbf{J}_t^{\transpose}\mathbf{H}^{\transpose}
\mathbf{w}s + z_{u},
\end{align}
where $\mathbf{q} \in \mathbb{C}^{M \times 1}$ comprises the $M$ diagonal entries of $\mathbf{\Phi}$ of the phase-shifts vector of the non-diagonal RIS. Accordingly, the received signal power at the user is obtained as follows:
\begin{align}\label{received_power}
P_r = P_t
|\mathbf{q}^{\transpose} 
\operatorname{diag}\{\mathbf{g}^{\transpose}\mathbf{J}_r^{\transpose}\} \mathbf{J}_t^{\transpose}\mathbf{H}^{\transpose}
\mathbf{w}|^2,
\end{align}
where $P_t$ is the BS transmit power.
It is evident from \eqref{received_power} that there is a need for the acquisition of the individual channels in order to optimize the switch array of the non-diagonal RIS. Specifically, this is necessary for designing $\mathbf{J}_t$ and $\mathbf{J}_r$ that maximize the received power $P_r$ of the downlink signal at the user.
Hence, in the next section, we provide a detailed description of the proposed channel estimation method, which facilitates the design of the optimal  $\mathbf{J}_t^{\kl}$ and $\mathbf{J}_r^{\kl}$.
\section{Proposed Three-Stage Channel
Estimation Framework}\label{sec:proposed_method}
A key challenge in non-diagonal RIS-assisted communication systems arises due to the lack of built-in signal processing capabilities or RF sources in RIS hardware, making it difficult to directly implement the optimal element mapping required for non-diagonal configurations. Specifically, achieving the optimal configuration requires the BS to perform exhaustive channel estimation across $M^2$ channel coefficients for a non-diagonal RIS with $M$ elements to identify the configuration that maximizes the channel gain. This approach incurs prohibitive overhead, severely degrading spectral efficiency, particularly in scenarios with limited coherence time. Addressing this challenge is essential for realizing the practical benefits of non-diagonal RIS architectures. To this end, we propose a three-stage channel estimation protocol that significantly reduces the overhead by requiring only three RIS configurations.
As shown in Fig. \ref{sysmo}, in the first two stages, the switch array is utilized to estimate scaled versions of  $\mathbf{H}$ and $\mathbf{g}$. These scaled estimates provide sufficient information to determine the optimal switch array configuration that maximizes the channel gains. Finally, in the third stage, the cascaded channel is estimated using the identified optimal configuration, enabling the joint optimization of active and passive beamforming.
We leverage the binary-reflection (ON/OFF) protocol \cite{on_off}, a widely adopted method for channel estimation in conventional RIS systems, where each of the $M$ elements is sequentially activated for a duration $\tau_m$.
Moreover, the  LS estimator is used due to its practicality and effectiveness in efficient wireless system design \cite{reasonable1}\footnote{If the direct channels between the BS and the user are non-negligible, their CSI can be obtained by switching off the RIS and then subtracted from the composite RIS channel observations, similar to \cite{on_off}.}.

\subsection{First Stage}
During the training period of the coherence interval, the user transmits a pilot signal $x_p$, having power $P_{\pil} = |x_p|^2$ to the BS. Thus, the received pilot signal at the BS is given by:
\begin{align}
\mathbf{Y}^{\ul} = \mathbf{F}^{\ul} \mathbf{X}_p + \mathbf{Z}_{\mathrm{BS}},
\end{align}
where $\mathbf{F}^{\ul}=\mathbf{H}\mathbf{J}_t^{\ul}\operatorname{diag}\{\mathbf{J}_r^{\ul}\mathbf{g}\} $, $\mathbf{X}_p \buildrel \Delta \over =x_p \otimes \mathbf{I}_M$ \cite{on_off}, and $\mathbf{Z}_{\mathrm{BS}}$ $\in \mathbb{C}^{N \times M}$ is the AWGN matrix at the BS with variance $\sigma^2_z$ for all entries.
To configure the non-diagonal RIS such that each antenna at the BS acquires a scaled version of the user-RIS channels i.e., the incident signals on the RIS elements are sequentially reflected back from the same element, one row of the permutation matrix $\mathbf{J}_t$ is set to ones and the other rows remain zeros. Meanwhile, $\mathbf{J}_r$ is set as an identity matrix. Hence, $\mathbf{J}_t$ and $\mathbf{J}_r$ are given as follows:
\begin{align}
\mathbf{J}_t^{\ul} \triangleq
\begin{bmatrix}
\mathbf{1}_{1 \times M} \\
\mathbf{0}_{(M-1) \times M}
\end{bmatrix} 
, \quad \mathbf{J}_r^{\ul} \triangleq \mathbf{I}_M,
\end{align}
where $\mathbf{1}_{1 \times M}$
is a vector with all elements equal to $1$, $\mathbf{0}_{(M-1) \times M}$ is a zero matrix with dimensions $(M-1) \times M$, and $\mathbf{I}_M$ is the Identity matrix with dimensions $M \times M$. 
Then, using the LS estimator, $\hat{\mathbf{F}}^{\ul}$ can be derived as:
\begin{align}
\hat{\mathbf{F}}^{\ul} = \mathbf{Y}^{\ul}\mathbf{X}_p^\dagger ,
\end{align}
where $\mathbf{X}_p^\dagger= \mathbf{X}^{\herme}_p  
\qty( \mathbf{X}_p \mathbf{X}^{\herme}_p )^{-1}$ is the pseudoinverse of $\mathbf{X}_p$. The LS estimation error is computed as:
\begin{align}
\tilde{\mathbf{F}}^{\ul} = \mathbf{Z}_{\mathrm{BS}}\mathbf{X}_p^\dagger.
\end{align}
Note that $\hat{\mathbf{F}}^{\ul}$ is a matrix with $N$ row vectors, i.e., $\hat{\mathbf{F}}^{\ul} = [\hat{\mathbf{f}}^{\ul}_1, \dots,\hat{\mathbf{f}}^{\ul}_n, \dots, \hat{\mathbf{f}}^{\ul}_N]^{\transpose}$, where each vector $\hat{\mathbf{f}}^{\ul}_n\in \mathbb{C}^{1 \times M}$ represents  a scaled version of  $\mathbf{g}$. To further enhance the channel estimation quality, we select the BS antenna with the highest average channel magnitude as the representative of the scaled version of $\mathbf{g}$. Specifically, let $\hat{\mathbf{f}}^{\ul}_{\mathrm{sel}}$  denote the selected vector, which is determined  based on the following criterion:
\begin{align}
\hat{\mathbf{f}}^{\ul}_{\mathrm{sel}} = \max\qty{ \frac{1}{M} \sum_{m=1}^M |\hat{\mathbf{f}}^{\ul}_n|}.
\end{align}
\subsection{Second Stage}
The second stage aims to configure the non-diagonal RIS switch array to enable the estimation of scaled versions of the RIS-BS channels. Hence, the switch array is configured such that one specific element of the RIS is connected to sequentially reflect signals from all other elements to the antennas of the BS. To achieve this configuration, the permutation matrix $\mathbf{J}_r$ is designed such that one column is set to ones while the remaining columns are zeros, as given by:
\begin{align}
\mathbf{J}_r^{\dl} \triangleq
\begin{bmatrix}
\mathbf{1}_{M \times 1} ,\mathbf{0}_{M \times (M-1)}
\end{bmatrix} 
, \quad \mathbf{J}_t^{\dl} \triangleq \mathbf{I}_M.
\end{align}
With this configuration, the received pilot signal can be expressed as:
\begin{align}
\mathbf{Y}^{\dl} = \mathbf{F}^{\dl} \mathbf{X}_p + \mathbf{Z}_{\mathrm{BS}},
\end{align}
where $\mathbf{F}^{\dl}=\mathbf{H}\mathbf{J}_t^{\dl}\operatorname{diag}\{\mathbf{J}_r^{\dl}\mathbf{g}\}$. By employing the LS estimator, the channel estimate $\hat{\mathbf{F}}^{\dl}$ is derived as:
\begin{align}
\hat{\mathbf{F}}^{\dl} = \mathbf{Y}^{\dl}\mathbf{X}_p^\dagger.
\end{align}
The corresponding LS estimation error $\tilde{\mathbf{F}}^{\dl}$ is computed as: 
\begin{align}
\tilde{\mathbf{F}}^{\dl} = \mathbf{Z}_{\mathrm{BS}}\mathbf{X}_p^\dagger.
\end{align}
\subsection{Third Stage}
In this stage, the BS utilizes the estimates $\hat{\mathbf{f}}^{\ul}_{\mathrm{sel}}$ and $\hat{\mathbf{F}}^{\dl}$, which represent scaled versions of the channels $\mathbf{g}$ and $\mathbf{H}$, respectively.  These estimates enable the identification of the optimal permutation matrices that configure the non-diagonal RIS for improved performance, as elaborated in the next section.
%
With this configuration in place, cascaded channel estimation is performed to acquire the complete CSI required for joint optimization of active and passive beamforming in the communication system. The received pilot signal can be expressed as:
\begin{align}
\mathbf{Y}^{\cas} = \mathbf{F}^{\cas} \mathbf{X}_p + \mathbf{Z}_{\mathrm{BS}},
\end{align}
where $\mathbf{F}^{\cas}=\mathbf{H}\mathbf{J}_t^{\kl}\operatorname{diag}\{\mathbf{J}_r^{\kl}\mathbf{g}\}$.
Therefore, employing the LS estimator, $\hat{\mathbf{F}}^{\cas}$ can be derived as:
\begin{align}
\hat{\mathbf{F}}^{\cas} = \mathbf{Y}^{\cas}\mathbf{X}_p^\dagger.
\end{align}
The corresponding  LS estimation error $\tilde{\mathbf{F}}^{\cas}$ is given by:
\begin{align}
\tilde{\mathbf{F}}^{\cas} = \mathbf{Z}_{\mathrm{BS}}\mathbf{X}_p^\dagger.
\end{align}
\section{Switch Array and Beamforming Design}
This section presents the design of the permutation matrices based on the scaled channel estimates obtained in the first and second stages, and the beamforming design based on the cascaded channel estimates of the third stage. We consider the SISO and MISO communication sytems.
\subsection{Case I: SISO System}
\subsubsection{Switch Array Design}
After the first and second stages, the BS has acquired the estimates $\hat{\mathbf{f}}^{\ul}$ and $\hat{\mathbf{f}}^{\dl}$,  which are the scaled versions of the channel vectors $\mathbf{g}$ and $\mathbf{h}$, respectively. For simplicity,
we denote $\hat{\mathbf{f}}^{\ul}=\mathbf{g}'$ and $\hat{\mathbf{f}}^{\dl}=\mathbf{h}'$. Hence, the maximum channel gain can be achieved when the amplitudes of the two vectors are sorted and multiplied element-wise \cite{non_diagonal}. For example, consider a RIS with $4$-elements, 
where the amplitudes of the scaled versions of the incident and reflected channels are sorted as follows:
\[g_{(1)}' = |g_4'|, \quad g_{(2)}' = |g_3'|, \quad g_{(3)}' = |g_1'|, \quad g_{(4)}' = |g_2'|,\]
\[
h_{(1)}' = |h_2'|, \quad h_{(2)}' = |h_3'|,\quad h_{(3)}' = |h_4'|,\quad h_{(4)}' = |h_1'|,\]
where $h_{(1)}'$, $h_{(2)}'$, $\dots$, $h_{(M)}'$ and $g_{(1)}'$, $g_{(2)}'$, $\dots$, $g_{(M)}'$ represent the sequence of sorted amplitudes in an ascending order for $\mathbf{h}'$ and $\mathbf{g}'$, respectively.
Hence, the optimal permutation matrices $\mathbf{J}_r^{\kl}$ and $\mathbf{J}_t^{\kl}$ are designed to realize this mapping as follows: 
\begin{equation}
\mathbf{J}_r^{\kl}
= 
\begin{bmatrix}
0 & 0 & 0 & 1 \\
0 & 0 & 1 & 0 \\
1 & 0 & 0 & 0 \\
0 & 1 & 0 & 0
\end{bmatrix},~~~~~~~
\mathbf{J}_t^{\kl}
= 
\begin{bmatrix}
0 & 0 & 0 & 1 \\
1 & 0 & 0 & 0 \\
0 & 1 & 0 & 0 \\
0 & 0 & 1 & 0
\end{bmatrix}.
\end{equation}
\subsubsection{Passive Beamforming Design}
After the third stage, the optimal phase-shifts vector $\mathbf{q}_{\text{opt}}$ is obtained from $\hat{\mathbf{f}}^{\cas}$ as:
\begin{align}\label{optimums}
\mathbf{q}_{\text{opt}} = \exp\qty{ -j  \angle\hat{\mathbf{f}}^{\cas}}.
\end{align}
\subsection{Case II: MISO system}
\subsubsection{Switch Array Design} In the MISO case, the BS acquires the estimates $\hat{\mathbf{f}}^{\ul}_{\mathrm{sel}}$ and $\hat{\mathbf{F}}^{\dl}$ during the first and second stages, which are the scaled versions of $\mathbf{g}$ and $\mathbf{H}$, respectively. we denote $\hat{\mathbf{f}}^{\ul}_{\mathrm{sel}}=\mathbf{g}'$ and $\hat{\mathbf{F}}^{\dl}=\mathbf{H}'$. The BS-RIS channel estimates is a matrix $\mathbf{H}' = [\mathbf{h}'_1, \mathbf{h}'_2, \dots, \mathbf{h}'_N]^{\transpose}$ with $N$ rows. In this case,  $\mathbf{J}_t$ cannot be designed in the same way as in the SISO scenario due to the additional complexity introduced by the multiple antennas.
To avoid the search complexity and efficiently utilize the obtained scaled versions $\mathbf{H}'$, a sub-optimal method is employed. For each RIS element, we compute the average channel magnitude across the $N$ antennas, i.e., $\bar{\mathbf{h}'} = \frac{1}{N} \sum_{n=1}^N |\mathbf{h}_n'|$. The resulting vector $\bar{\mathbf{h}'}$ $\in \mathbb{C}^{1 \times M}$ is a row vector, and thus, the permutation matrices $\mathbf{J}_t$ and $\mathbf{J}_r$ are designed to sort the amplitudes of the row vector $\bar{\mathbf{h}'}$ and the column vector $\mathbf{g}'$, respectively, in ascending order \cite{non_diagonal}. \vspace{1mm}
\subsubsection{Active and Passive Beamforming Design}
Recall that, the permutation matrices $\mathbf{J}_t$ and $\mathbf{J}_r$ are designed after the first and second stages of the proposed three-stage protocol which in turn are utilized in the third stage to acquire the cascaded channel estimates $\hat{\mathbf{F}}^{\cas}$. This significantly simplifies the active and passive beamforming optimization for the downlink, as the complexity associated with searching for these matrices during the optimization process is eliminated. Utilizing \eqref{received_power}, we employ an alternating optimization algorithm to maximize the spectral efficiency as follows:
\begin{equation}
\begin{aligned}
& \underset{\mathbf{w},\, \mathbf{q}}{\text{maximize}} 
&& \scalemath{0.9}{\qty(1-\frac{3M\tau_m}{\tau})}
\scalemath{0.9}{\log_2\qty(1+ \frac{P_t}{\sigma^2_z}\qty| \mathbf{q}^{\transpose}\mkern2mu \hat{\mathbf{F}}^{{\cas}^{\transpose}}\mkern2mu
\mathbf{w} |^2)} \\
& \text{subject to} && \|\mathbf{w}\| = 1, \\
& && \qty|\qty[\mathbf{q}]_{m}|=1,\quad m = 1, 2, \ldots, M.
\end{aligned}
\end{equation}

\begin{algorithm}
\caption{Alternating Optimization for Downlink Active and Passive Beamforming}
\label{alg}
\begin{algorithmic}[1]
\State \textbf{Input:} Third stage cascaded channel estimate $\hat{\mathbf{F}}^{\cas}$.
\State \textbf{Output:} Optimal phase-shifts vector  $\mathbf{q}_{\text{opt}}$ and optimal beamforming vector $\mathbf{w}_{\text{opt}}$.
\State Initialize random $\mathbf{w}$ such that $\|\mathbf{w}\| = 1$.
\Repeat
    
    \State Design the optimal phase-shifts vector $\mathbf{q}$ as:
    \[\mathbf{q} = \exp\qty{ -j  \angle\qty(\hat{\mathbf{F}}^{{\cas}^{\transpose}}\mathbf{w})} 
\]
    \State Update the BS beamforming vector $\mathbf{w}$ based on the maximum ratio transmission (MRT) scheme:
    \[
        \mathbf{w} = \frac{\qty(\mathbf{q}^{\transpose}\hat{\mathbf{F}}^{{\cas}^{\transpose}})^{\herme}}{\big\| \mathbf{q}^{\transpose}\hat{\mathbf{F}}^{{\cas}^{\transpose}}\big\|}
    \]
\Until{maximum number of iterations reached or convergence.}
\State Set $\mathbf{q}_{\text{opt}}=\mathbf{q}$
and
$\mathbf{w}_{\text{opt}} = \frac{\qty(\mathbf{q}^{\transpose}_{\text{opt}}\hat{\mathbf{F}}^{{\cas}^{\transpose}})^{\herme}}{\big\| \mathbf{q}^{\transpose}_{\text{opt}}
\hat{\mathbf{F}}^{{\cas}^{\transpose}} \big\|}$.
\end{algorithmic}
\end{algorithm}
Note that, Algorithm \ref{alg} can also be applied in the case of the conventional diagonal RIS with the difference that the permutation matrices are set as identity matrices $(\mathbf{J}_t=\mathbf{J}_r=\mathbf{I}_M)$, and the channel estimation time is $M\tau_m$.
\section{Complexity Analysis}
In this section, we assess the complexity of the estimation process for the proposed three-stage protocol compared to exhaustive estimation. In the case of exhaustive estimation, the BS requires to estimate the CSI of $M^2$ coefficients of the $M \times M$ switch array, resulting in a computational complexity of $\mathcal{O}\qty(N M^2)$.
In contrast, our proposed scheme significantly reduces this complexity by requiring only $3M$ coefficients to be estimated for each of the $N$ antennas. Thus, the overall computational complexity of our approach is $\mathcal{O}\qty(NM)$, representing a substantial reduction in computation time and resource usage.
Note that this complexity is similar to the conventional ON/OFF-based LS estimation for diagonal RIS, where only a single stage is needed to estimate the $M$ cascaded channels, resulting in an estimation time of $M\tau_m$.
\section{Numerical Results}\label{sec:numerical}
In this section, we evaluate the proposed channel estimation scheme. We randomly generate $10^5$ Rayleigh fading channels and execute the entire channel estimation procedure for each realization. The distances between the BS and the non-diagonal RIS is $d_{\RIS}=30$ m, and the distance between the non-diagonal RIS to the user $d_{\RISg}=20$ m. The path-loss exponents are set $\eta_H=\eta_g=3$, and the noise power is $\sigma_{z}^2=-80$ dBm.

\begin{figure}
\centering
\begin{multicols}{3}
\includegraphics[width=1.15in]{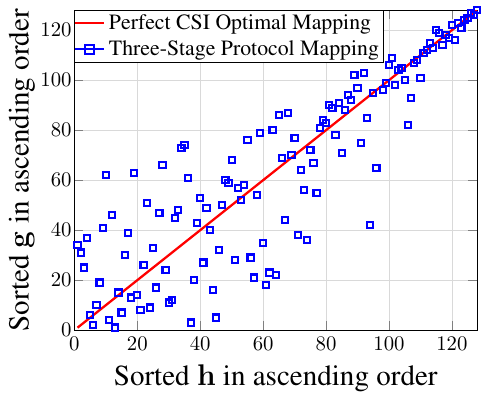}
\vspace{-5.75mm}
\renewcommand{\figurename}{\scriptsize Fig.}
\subcaption{\scriptsize  $P_u=-15$ dBm.}
\label{1w_ref}
\par \hspace*{-2mm} 
\includegraphics[width=1.15in]{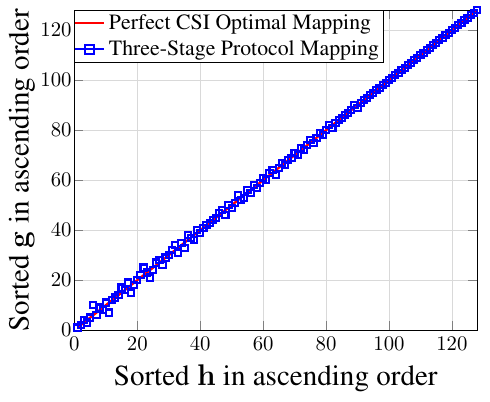}
\vspace{-5.75mm}
\renewcommand{\figurename}{\scriptsize Fig.}
\subcaption{\scriptsize  $P_u=0$ dBm.}
\label{slight_mis}
\par \hspace*{-3mm}$\mkern-3mu$ 
\includegraphics[width=1.15in]{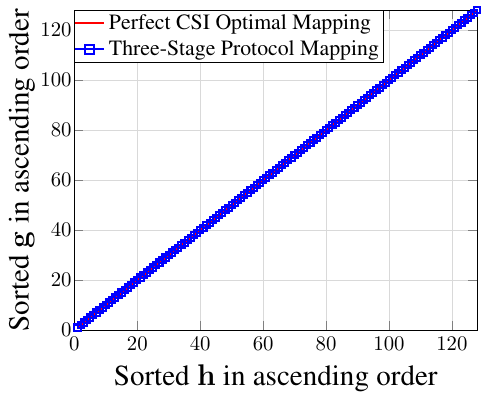}
\vspace{-5.75mm}
\renewcommand{\figurename}{\scriptsize Fig.}
\subcaption{\scriptsize  $P_u=15$ dBm.}
\label{3w_ref}
\par 
\end{multicols}
    \renewcommand{\figurename}{\footnotesize Fig}
    \vspace{-9mm}
\caption{\footnotesize Mapping of elements of the SISO case for different $P_u$ with $M=128$.}
\vspace{0mm}
      \label{three_fig}
\end{figure}

In Fig \ref{three_fig}, the mapping accuracy of elements via the non-diagonal RIS switch array is shown for different uplink power levels $P_u$ with $M=128$ reconfigurable elements. The results demonstrate that as the uplink power increases, the mapping accuracy improves. This is because the optimization of the switch array connections is integral to the channel estimation process, which relies on accurately estimating the scaled versions of the incident and reflected channels during the first and second stages.  Specifically,  at very low power values, the mismatch is more random, while at higher powers, mismatches tend to occur between elements with similar channel strengths. As illustrated in Fig. \ref{slight_mis}, mismatches involve swapping values that are close in magnitude within the sorted sequence. This suggests a promising compromise between the uplink power and mapping accuracy, as such comparable mismatches can have a negligible effect on the maximum channel gain.
%
%
%
%
%
%
%
%
%
%
%
%
\begin{figure}[t]
    \centering
    \includegraphics[width=2.5in]{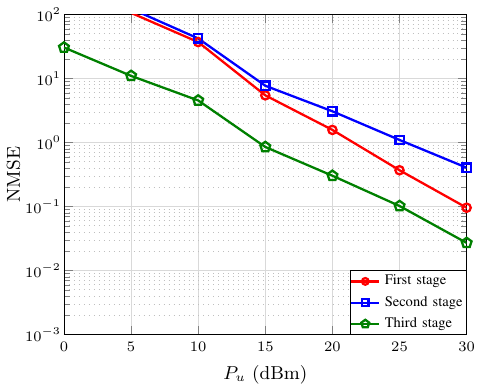}
\vspace{-2.0mm}
\renewcommand{\figurename}{\footnotesize Fig.}
\captionwidth
\caption{\footnotesize NMSE performance of the each stage for $N=4$ and $M=64$.}
\label{NMSE}
\vspace{-4mm}
\end{figure}
\begin{figure}[t]
    \centering
    \includegraphics[width=2.5in]{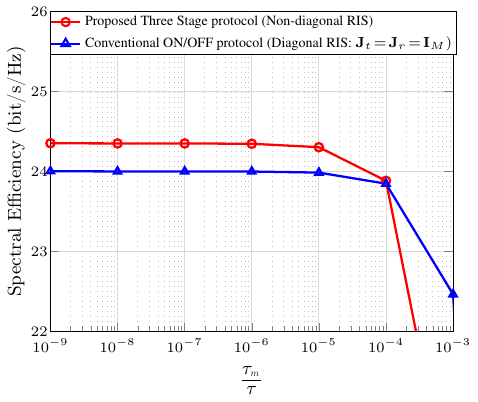}
\vspace{-2.0mm}
\renewcommand{\figurename}{\footnotesize Fig.}
\captionwidth
\caption{\footnotesize  The Spectral efficiency versus the ratio $\frac{\tau_{m_{}}}{\tau}$ for $N=4$ antennas and $M=64$ elements, $P_u=30$ dBm and BS transmit power $P_t=40$ dBm (downlink power is 10 times greater than the uplink pilot power \cite{two_time}).}
\label{coh_time}
\vspace{-3mm}
\end{figure}

In Fig. \ref{NMSE}, we present the normalized mean square error (NMSE) across all three stages. The NMSE can be  defined as follows \cite{two_time}:
$
\mathrm{NMSE} = \frac{\e\qty{\|\hat{\mathbf{A}} - \mathbf{A}\|_2^2}}{\e\qty{\|\mathbf{A}\|_2^2}},
$
where $\hat{A} \in \qty{\hat{\mathbf{f}}^{\ul}, \hat{\mathbf{F}}^{\dl}, \hat{\mathbf{F}}^{\cas}}$ and $A \in \qty{\mathbf{f}^{\ul}, \mathbf{F}^{\dl}, \mathbf{F}^{\cas}}$. One can observe that the third stage achieves the lowest NMSE among the three stages. This is expected as the third stage fully exploits the degrees of freedom of the $M$ elements ($M$ degrees of freedom) to obtain the cascaded channels even if the sorting is still random. In contrast, the first and second stage only exploit one element to be connected to the other channels to obtain the scaled estimates, and hence, only one degree of freedom is achieved. It is also observed that, the first stage achieves better NMSE performance than the second stage, due to the utilization of the BS antennas degrees of freedom, where the BS selects the antenna with the highest average channel magnitude to represent the scaled estimates of the user–RIS channels.

\begin{figure}[t]
    \centering\hspace{-5mm}
    \includegraphics[width=2.6in]{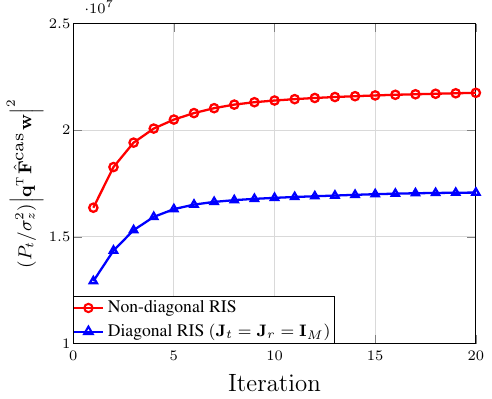}
\vspace{-2.0mm}
\renewcommand{\figurename}{\footnotesize Fig.}
\captionwidth
\caption{\footnotesize Convergence of Algorithm $1$ for non-diagonal and diagonal RIS, $P_u=30$ dBm,
$P_t=40$ dBm, the curves are averaged over 100 Monte-Carlo run. }
\label{conve}
\vspace{-4mm}
\end{figure}
Fig. \ref{coh_time}  illustrates the comparison of downlink spectral efficiency achieved using the proposed three-stage protocol and the conventional estimation method for the diagonal RIS as a function of the ratio $\frac{\tau_{m_{}}}{\tau}$,
offering a benchmark for comparison against the proposed three-stage protocol. To enable a fair comparison, we also consider the conventional ON/OFF protocol combined with LS channel estimation.
It is observed that the proposed three-stage estimation protocol can realize the non-diagonal configuration, and thus, higher spectral efficiency is achieved compared to the conventional diagonal RIS. However, its performance degrades at higher values of $\frac{\tau_m}{\tau}$ compared to the conventional approach. This is attributed to the fact that the time required for channel estimation for the three-stage protocol is $3M\tau_m$, while the conventional approach for diagonal RIS requires $M\tau_m$. Consequently, the conventional diagonal RIS is expected to offer higher spectral efficiency in high-mobility scenarios.

Fig. \ref{conve} evaluates the convergence rate of the proposed alternating optimization algorithm. To provide also a comparison with the diagonal RIS, algorithm 1 is applied for diagonal RIS by setting the permutation matrices as identity ones, which reduces the non-diagonal RIS configuration to a diagonal one.  We conduct $100$ independent trials using different random initializations for the active beamforming vector $\mathbf{w}$.  It is observed that the algorithm converges for both non-diagonal and diagonal RIS nearly at the same number of iterations thanks to the proposed three staged protocol, which enabled the acquisition of the permutation matrices in the channel estimation process. As a result, the optimization of active and passive beamforming is significantly simplified, as there is no need to search for the optimal permutation matrices that maximize the channel gain.

\section{Conclusions}\label{sec:conc}
In this paper, we developed a novel three-stage channel estimation protocol for non-diagonal RIS-assisted communication systems, considering both SISO and MISO configurations. Our approach leverages the switch array of the non-diagonal RIS during the channel estimation process, enabling the acquisition of scaled versions of the user-RIS and RIS-BS channels in the first and second stages. This allows us to optimize the element mapping of the switch array for maximum channel gains. In the third stage, the cascaded user-RIS-BS channels are estimated to enable the optimization of active and passive beamforming. The proposed scheme significantly reduces the computational complexity at the BS from $\mathcal{O}\qty(N M^2)$ to $\mathcal{O}\qty(NM)$, which is similar to the complexity of ON/OFF-based LS estimation for conventional diagonal RIS and eliminates the need for active sources at the RIS. These advantages make the protocol highly efficient and practical for real-world implementation in non-diagonal RIS-assisted systems.

\vspace{0mm}

\linespread{1.1}
\vspace{-3mm}
\sloppy
\bibliographystyle{IEEEtran}
\bibliography{IEEEabrv,References}
\pagestyle{empty}

\end{document}